# Design Differences between the Pan-STARRS PS1 and PS2 Telescopes

Jeffrey S. Morgan*[a], Nicholas Kaiser[a], Vincent Moreau[b], David Anderson[c], and William Burgett[a]

[a]Institute for Astronomy, Pan-STARRS MIC Suite 198, 2680 Woodlawn Dr., Honolulu, HI, USA 96822; [b]AMOS, Liege Science Park, Rue des Chasseurs Ardennais, B-4031 Angleur (Liege), Belgium; [c]Rayleigh Optical Corp., 3720 Commerce Dr., Suite 1112, Baltimore, MD, USA 21227;

## ABSTRACT

The PS2 telescope is the second in an array of wide-field telescopes that is being built for the Panoramic-Survey Telescope and Rapid Response System (Pan-STARRS) on Haleakala. The PS2 design has evolved incrementally based on lessons learned from PS1, but these changes should result in significant improvements in image quality, tracking performance in windy conditions, and reductions in scattered light. The optics for this telescope are finished save for their coatings and the fabrication for the telescope structure itself is well on the way towards completion and installation on-site late this year (2012). The most significant differences between the two telescopes include the following: secondary mirror support changes, improvements in the optical polishing, changes in the optical coatings to improve throughput and decrease ghosting, removal of heat sources inside the mirror cell, expansion of the primary mirror figure control system, changes in the baffle designs, and an improved cable wrap design. This paper gives a description of each of these design changes and discusses the motivations for making them.

## 1. INTRODUCTION

The Pan-STARRS Project is an effort to field an array of 2-m class telescopes which will work in unison to provide rapid, nearly full-sky coverage in 5 colors. The first telescope/camera system for this project, called PS1, has already been built and has been in constant operation since 2010. The synoptic data gathered from PS1 has been proven to be useful for the study of temporal variations in the sky including moving objects such as asteroids as well as for the development of one of the most accurate photometric and astrometric all-sky catalogs available to date. The second telescope, PS2, is currently in fabrication and is expected to come on-line in 2014. There were several lessons learned in the development of PS1 which we are attempting to implement in the fabrication of PS2. This paper is a description of the differences between the design and fabrication details of PS2 from the original, prototype telescope.

The differences between the design and fabrication of PS2 and PS1 include the following items:

1. A change in the design of the secondary (M2) support structure.
2. Improved polishing of all of the optics (in particular, the L2 corrector)
3. Changes in the anti-reflection (AR) coatings of the corrector optics.
4. Removal of heat sources inside the primary (M1) mirror cell.
5. Improvements in the cable wrap design.
6. Changes in the M1 figure control actuators.
7. Improved baffling.

We will describe the details of all of these changes in the rest of this paper. We believe that the incremental improvements described here will offer substantial gains in the performance of the PS2 telescope over that currently seen in PS1. For a description of the PS2 camera one should see the article at this conference by Onaka, et al[1]. And for a description of schedule and management of the PS2 fabrication one should read the article by Burgett[2].

## 2. THE M2 SUPPORT STRUCTURE

Figure 1 shows a photograph of the PS1 support structure. The axial support structure is a Hindle type 18-point whiffletree mount utilizing three bi-pods and six tripods that attach the secondary to the moving end of a PI hexapod. The lateral support is a passive counterweight design. Twelve of the 18 counterweights may be seen as steel cylinders that extend behind and around the perimeter of the M2 mirror. The original design of the support structure utilized 2-dimensional perforated hinge flexures at every whiffletree support point. This design was very forgiving of assembly errors, but it was rapidly seen that this structure was too flexible and was very susceptible to wind shake. Rubber dampers were placed around the counter weights which helped to minimize amount of time it took to settle out both slew-induced and wind-induced vibrations, but the structure was still too flexible to operate in winds higher than about 5 m/s. After about 9 months of operation with the original flexures, one of the axial flexures broke. It was discovered that there were design flaws which allowed the axial flexures to be over-stressed during assembly and that there were fabrication errors in the flexures themselves. The most critical fabrication error was found to be the failure to have had the steel in the flexures properly heat-treated. This resulted in a drastically reduced flexure lifetime and ultimately, the on-sky failure of one of the flexures. Because of that critical flaw, all of the support structure flexures had to be replaced and we took that opportunity to re-design the flexure arrangement in order to stiffen up the structure. The current PS1 secondary support flexure design utilizes 1-dimensional flexures that have been oriented in a manner so as to minimize the impact of thermal stress on the mirror. The current PS1 M2 support flexures are significantly stiffer than the original design and allow the telescope to operate in winds close to 10 m/s, but they are still marginal in performance. For this reason, we were very interested in using a significantly different and stiffer M2 support design for PS2.

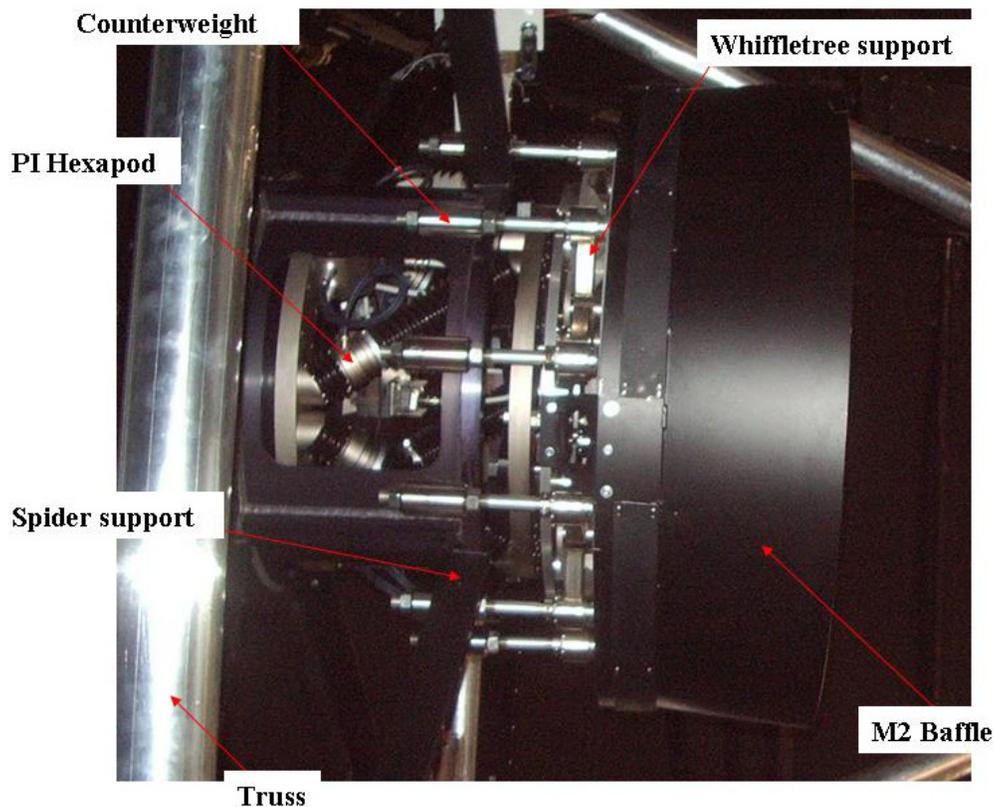

Figure 1. The PS1 secondary support

The PS2 M2 support was developed by the design team at the Advanced Mechanical and Optical Systems, Ltd. (AMOS) factory, the manufacturer chosen for the fabrication of the second telescope. This design is shown in Figure 2. The pink structure attaches to a hexapod which will be responsible for the movement of the secondary mirror. The design of the M2 mirror has not been altered from that used in the PS1 telescope. The 18-point axial whiffletree support structure has been maintained. Needle flexures now take the place of the perforated hinges. The needles are very similar in concept and performance to the original PS1 2-dimensional perforated hinges. However, in this M2 support design, the axial flexures are not required to simultaneously support the lateral loads from the mirror. In the new design, the lateral loads are carried by a post connected to the mirror by a thermally relieved central blade which is mounted in the center hole of the M2 mirror. The tripods in this design are significantly thicker than for those of the PS1 design owing to the need to support the longer needle flexures that are being employed by this structure.

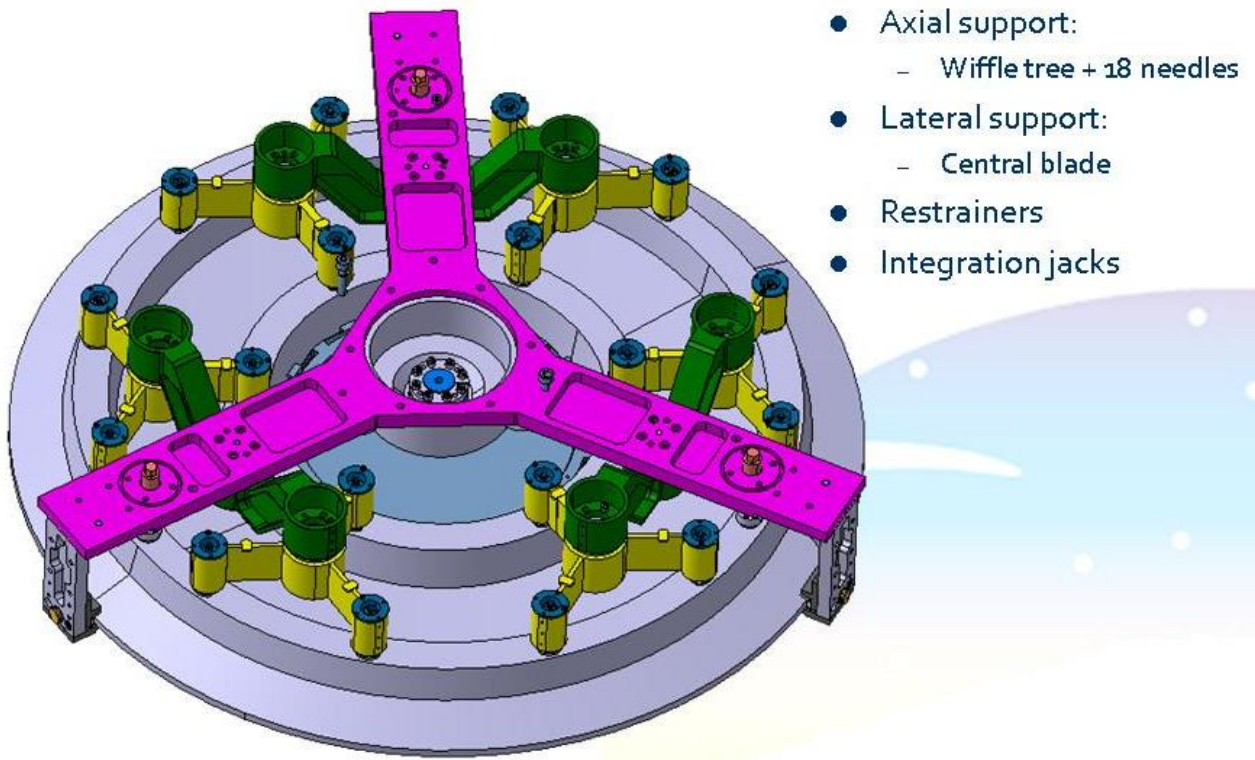

Figure 2. The PS2 secondary support (figure courtesy of AMOS)

The new M2 support structure is significantly stiffer than the PS1 design and we anticipate that this will help significantly in our ability to observe in windy conditions. It has a first resonant frequency of ~40 Hz, compared to the modified PS1 design which has its first resonant frequency at ~10 Hz.

## 3. THE POLISHING OF THE PS1 & PS2 OPTICS

The designs of the PS2 and PS1 optics are identical. Figure 3 shows that this layout consists of three refractive corrector lenses (L1-L3) along with the primary (M1) and secondary (M2) mirrors. Six filters are available in the PS1/2 telescopes. They are inserted into the optical path at the three different positions (F1-F3) shown in this layout. Owing to the nature of the survey's scientific goals, the passband of the system is fairly large, going from 395 nm to 1100 nm. The filters used have passbands of ~100 nm or more. In order of increasing central wavelengths, the filter passbands are designated as g, r, i, z, y, and w. The g, r, i, and z passbands are very close to those of the Sloan Digital Sky Survey (SDSS) of the same names. The w filter is a specialized wideband filter used for asteroid surveying that extends over the combined passbands of the g, r, and i filters. Only one filter at a time is used in this system.

This figure shows that the L2 corrector lens is both very thin (~13 mm thick at its vertex) and very deep. Both surfaces of this corrector lens are spherical, while each of the other optics employ aspheric surfaces. The corrector diameters are 0.61, 0.57, and 0.55 m for L1, L2, and L3, respectively.

Improvements have been made in the polishing of all of the optics that will go into the PS2 telescope. We believe that the most significant change in the optics will come from better figuring of the PS2 L2 corrector lens. However, we also believe that improvements in the polishing of the L3 corrector lens will also have a significant impact.

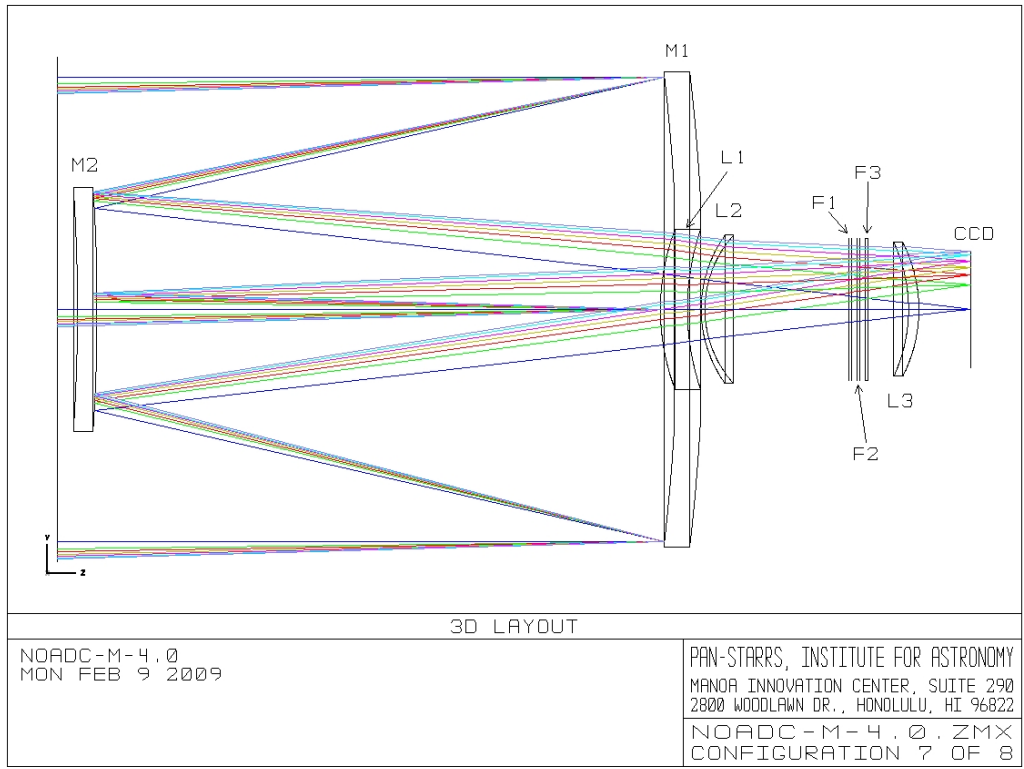

Figure 3. The optical layout of the PS2 telescope.

### 3.1 The PS1 Optics

We currently believe that the optical quality of the PS1 telescope is limited mostly by errors in the polishing of its L2 corrector lens, with a contribution at the center of the optical axis coming from the polishing of the L3 lens. During initial fabrication, the rough figure generation of the L2 lens was improperly done. Very significant high frequency scalloping in the lens substrate was placed on the glass which could not be corrected by polishing. The polishing vendor (OpticTechnium) was not the same as the vendor that did the L2 lens generation. The rest of the PS1 telescope optics, save the filters, were fabricated at Rayleigh Optical Corporation (ROC). The PS1 filters were manufactured by Barr Associates (now a subsidiary of Materion).

Through heroic efforts by the team at OpticTechnium, the L2 lens was saved by being reground to a slightly thinner configuration, but at some costs. The polishing vendor was not capable of doing full aperture testing of the lens and the delays caused by the regrinding prohibited us sending this lens out for such tests. All of these circumstances give us reasons to suspect the performance of the PS1 L2 corrector lens, but below we discuss other independent indications that this optic is the major limitation in the PS1 telescope performance.

Our initial investigation into the optics in the PS1 telescope failed to pinpoint a cause for the optical performance which yielded spot sizes ~2 times larger than the design expectations. This analysis included wavefront measurements of the system aberrations through analysis of out of focus donuts, a ray trace analysis of the behavior of aberrations based on glass positioning errors, and a simulation of the expected optical performance based on the measured full aperture polishing errors of all 5 of the other optics in the system.

The derivation of the wavefront errors in the system from measurements of out of focus donuts and on-sky measurements of stellar Point Spread Functions (PSFs) with co-incident seeing measurements were consistent with the PS1 optics producing a median value of ~0.5" FWHM averaged across the entire field of view. This performance was worse than what we were hoping to achieve for PS1, which was 0.41" median across the field of view. Given the seeing on Haleakala, this degradation is estimated to have an impact on the system approximately 10-15% of the time and has limited our best image quality achieved to date at the site of 0.63" FWHM averaged across the entire field of view, which includes degradations of the image quality from the camera, dome seeing, site seeing, and tracking errors. It is important to note that it is well established that with the exception of a small region close to the image center, the image quality in the PS1 telescope is largely independent of position in the FOV.

The ray trace analysis of the stack up of polishing errors in the rest of the optics (M1, M2, L1, L3, and the filters) in the system indicated that these optics were contributing at worst 0.36" FWHM to the system image quality. This was roughly within specification. However, there are reasons to suspect that there are errors in the analysis of the L3 interferometry near the very center of the lens. By process of elimination, it appears that the L2 optic is the main optical limitation. We show below that this is an entirely reasonable expectation.

Figure 4 shows the measured PS1 focal plane distortions. It is important to keep in mind that this graph does not show the image blur as a function of position in the focal plane, but rather it shows the position along the optical axis of best focus as a function of position in the focal plane. This is all relative to the best average focus position for the whole image. In effect, this shows the warping of the surface of best focus. The left side of this figure (4a) is a two-dimensional map of the height of the best focal surface. This particular figure was derived from an analysis of thousands of out of focus donuts distributed across the focal plane, but we have also derived essentially the same figure from local fits to stellar PSFs as a function of variations in the telescope focus. The X and Y axes in this graph are the position on the focal plane in meters. The square tiling observed in this figure is the result of small residual errors in the placement of the focal plane detectors. The grayscale bar on the right of 4a is scaled in units of microns. The circular symmetry present in this image is a very strong indication that the focal plane distortions come from the telescope optics. At the center of this image, the data saturate the chosen grayscale range.

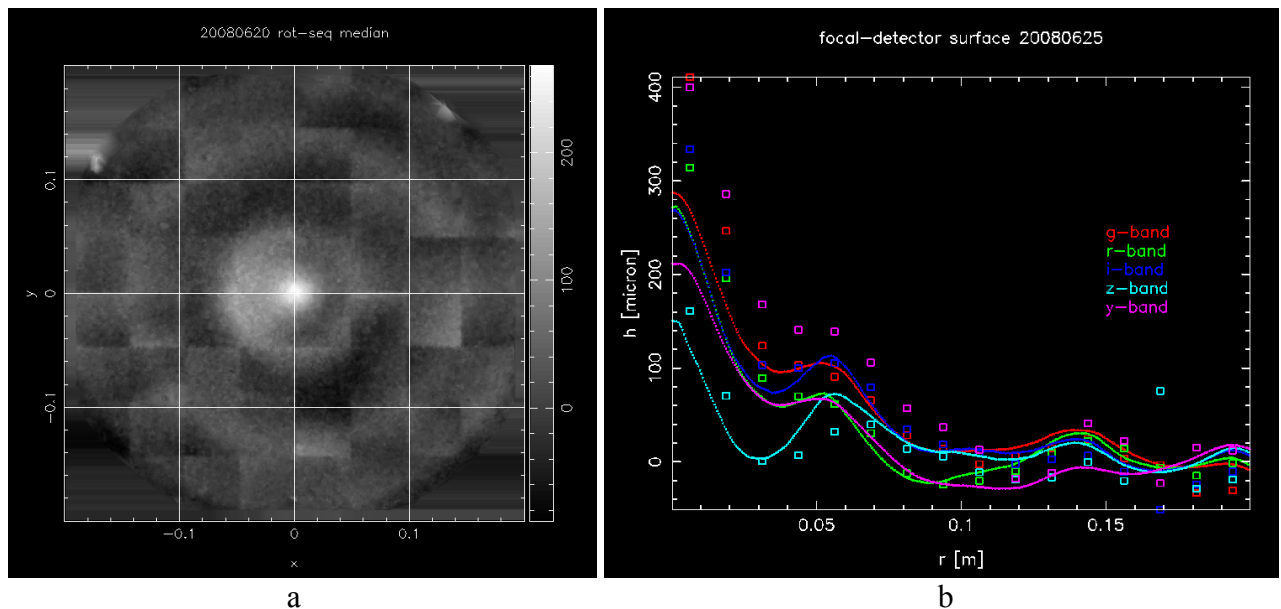

a                                                                                          b

Figure 4. Measured focal plane distortions in the PS1 telescope.

The graph shown in 4b is the azimuthal average of the distortions seen in 4a. In addition, this graph shows data taken through five of the filters in the system and it shows the results of two independent analyses. The symbols show the focal plane distortions measured by PSF fits, the lines show the focal plane distortions measured by an analysis of out of focus donuts. One can see the rings that are present on the left image along with a more quantitative measure of the height of the central peak distortion. There appears to be some filter dependence to the magnitude of the central peak. There is also a single odd outlying z-band point at a radius of 0.17 m. With those exceptions, the magnitude and locations of the rings at radii greater than 0.05 m are fairly consistent and show little or no filter dependence.

Rings of focal plane distortion are expected if the system is well collimated. And with the exception of the spike near the center of the image, the observed rings are of the expected magnitude. This is illustrated below in Figure 5. The radii of the outer rings can be made to vary with modest surface polish errors on several surfaces, so the locations of these rings are not very diagnostic. The presence of circular rings overlying a flat focal plane is important. This rules out almost all combinations of uncompensated tip, tilt, decenter, and spacing errors. There are several tilt errors that are degenerate with compensating decenter errors in this system. There are also some compensating despace errors. These modes can combine to restore circular symmetry in the system. However, these compensations also restore the independence of spot size as a function of position in the system, so they are not of interest in this analysis. Nor do we care about them in a practical sense. Uncompensated tip, tilt, and despace errors of the corrector lenses cause elliptical curvature to the focal plane surface and resultant variations in the system image quality with radius. Since the observed rings are symmetric, we can rule out these sorts of collimation errors in the system. This conclusion is strongly supported by an analysis of the positions of ghost images in the system[3]. This analysis showed that there is an uncompensated Y-axis tilt in the UCC corrector lenses of approximately -0.01°. This is at least an order of magnitude too small to have a significant impact on the PS1 image quality.

Figure 5 shows the focal plane distortions expected from the optical design in the system as a function of filter passband. The convention of relative offsets in this figure are inverted from that shown in Figure 4, so negative best focus positions in this graph correspond to positive offsets shown above. One can see from a comparison of the data in Figure 4 and Figure 5 that the amplitude of the measured central spike in PS1 is two to three times larger than expected. The measured central spike is also much narrower than expected. The central peak in the telescope focal plane appears to extend to a radius of only 0.02 to 0.03 m (0.14-0.21 degrees). This is important because the only optical surfaces capable of generating features this narrow are those on the L3 corrector. The central bulge in the design data extends out to a radius of 0.06 m. This size is characteristic of how fast the beam footprint changes the surface on L2 that it samples. It is easily verified through ray tracing that this central bulge in the design data is defined mostly by the shape of the L2 corrector lens surfaces. The radial position of the outer ring in the measured data is at 0.14 m. This is not well reproduced in the design data which shows an outer ring at 0.19 m. The design data also indicates that there should be a third ring out at the perimeter of the telescope FOV, somewhere around 0.25 m. There are some indications of this ring in the PS1 measurements shown in Figure 4, but the PS1 as-built measurements do not extend out far enough to make a firm conclusion on the existence of this ring.

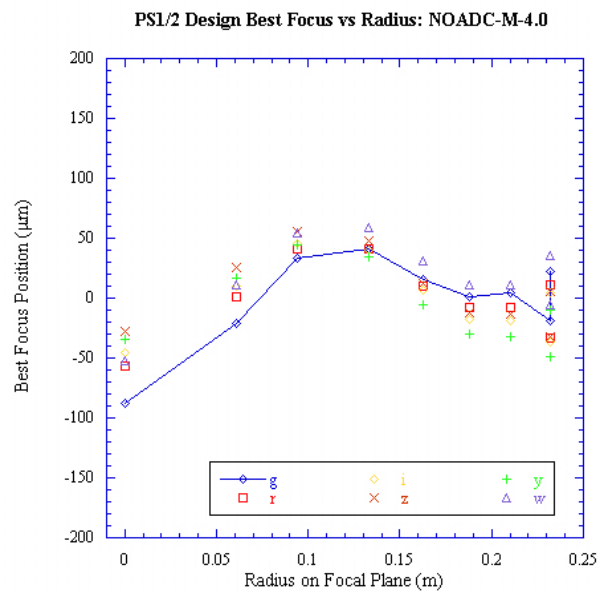

Figure 5. Expected focal plane distortions in the PS1/2 telescopes.

The design has some expected filter dependence. The g filter passband is expected to produce the largest central bulge and have very little ripple at large radii. The y filter passband is expected to have the opposite behavior. The observed central spike does not follow this trend. The largest amplitude central spikes are seen in both the g and y filters. The amplitude of the central spike does not follow any monotonic pattern with wavelength. The z passband, which is very close to the y passband, shows almost no spike. One might guess that this erratic behavior is therefore the result of polishing errors on the filter substrates, but this is inconsistent with the sharpness of the central spike. Surface errors on the filters are capable of adding to a central bulge, which might act as a broader pedestal that should extend out to ~0.05 m on which a spike resides, but they cannot cause the spike itself. The y passband data might be consistent with such an interpretation since it shows elevated focal surface distortions all the way out to a radius of 0.1 m. We also have data which show that the central spike remains without any filters in the beam. It has been for these reasons that we have tightened up the polishing specifications for the PS2 filter substrates and are now requiring full aperture interferometry of the filter substrate polish.

A critical point that is not illustrated by the data shown in Figure 4 is that for PS1 the spot size at minimum focus across the whole field of view is considerably elevated above the design spot size. The measured average PSF of 0.67" FWHM corresponds to an RMS spot size of 18 μm. It is well known from thousands of measurements that beyond the radius of the central spike (0.03 m < r), there is very little focal plane dependence to the telescope image quality. Typically there is less than a 0.1" difference in the FWHM measured at different regions on the focal plane.

The comparisons above between the PS1 performance and the design performance indicate that the problem with the PS1 optics is not a design issue, but a fabrication issue. The circular symmetry of the focal plane distortions and ray tracing calculations suggest that this is not a collimation problem either. Figure 6 shows that it is possible to explain a uniform degradation of the focal plane image quality with polishing errors on the L2 corrector.

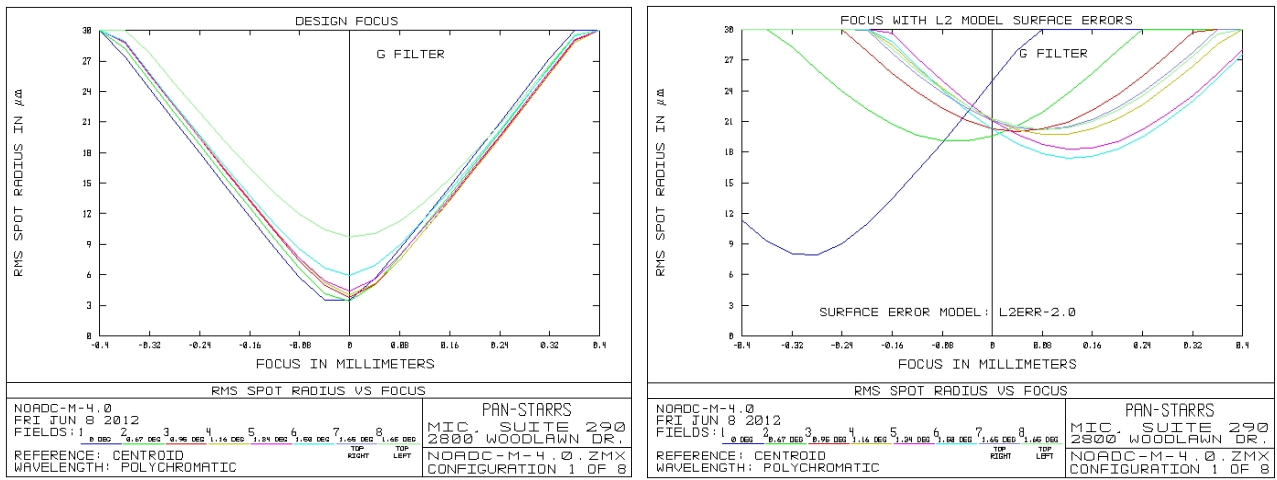

Figure 6. Expected focus vs radius in the PS1/2 telescopes.

Figure 6 shows a comparison of the telescope focus as a function of position in the field of view between the design (on the left) and the design with surface errors added only to the L2 corrector lens (on the right). The left side is indicative of what the design should provide, the right side is close to what the PS1 optics are currently providing. The colored lines in each graph show focus curves at different positions in the focal plane. The blue line is the on-axis focus curve. The other curves represent a sequence of increasing radii from the optical axis corresponding to the radii 0.094, 0.133, 0.163, 0.188, 0.210, and 0.232 m. There are two curves at the 0.232 m radius, one from the upper left of the image plane and the other from the upper right. The legend in each graph shows which color corresponds with which position, but the units in the legend is degrees from the optical axis. The black vertical line shown on each graph indicates the best focus compromise over the entire focal plane field of view. The L2 surface error model shown on the right side of this figure was originally constructed in an attempt to match the current PS1 image characteristics including the central spike. The shift of the on-axis focus curve (seen in blue) is a good mimic of the magnitude of the focal plane distortions seen at the center of the focal plane in PS1. However, these efforts served to illustrate that while it's possible to produce a central bulge of the right magnitude simultaneously with a uniform degradation of the telescope image quality, it is not possible to match the narrow shape of the central spike with only L2 surface errors.

The narrow width of the central peak implies a significant central surface error on L3. However, the calculations above show that at least half of the central peak amplitude can be explained by a broader contribution that could come from surface errors on L2. We will show below that the interferogram of L3 showed indications of on-axis surface errors, but these errors did not at first appear to be of sufficient magnitude to explain the observed distortions. However, we now believe that real on-axis surface errors on L3 were obscured in the interferogram by a diffraction spot from the hologram used to generate the interference pattern.

### 3.2 The PS2 Optics

The PS1 image quality degradations cannot be attributed to a single optic. However, it does appear that most of the problems come from a combination of L2 and L3 polishing errors with a possible minor contribution coming from the filter substrate polish. For these reasons, we have focused on improvements in the polishing of the L2 and L3 corrector optics. In addition to these improvements, we have also obtained better polish of all the other optics in the system.

Figure 7 compares the raw interferometry made on the PS1 and PS2 L3 corrector lenses. All of the corrector lenses were tested interferometrically in double pass transmission with a computer-generated hologram. On the left is the PS1 L3 interferogram and on the right is the PS2 L3 interferogram. The red circle on the PS1 L3 interferogram denotes the size of the optical beam footprint on the L3 corrector lens. This circle illustrates the region over which the on-axis beam samples the L3 surface. The noisy bright spot near the center of the PS2 interferogram is due to out-of-focus higher-order diffraction from the hologram. This spot is present in both the PS1 and the PS2 L3 interferograms, but it is much less obvious in the PS1 data. For both interferograms, the presence of the higher-order diffraction spot makes it difficult for the software to properly reconstruct the surface errors underneath the diffraction spot. With both PS1 & PS2 data, the on-axis data are masked from the reconstructive software analysis. With the PS1 L3 lens, the contours leading up to this central region indicate that there is significant on-axis structure that should not have been disregarded. In contrast, one can easily see that there is no such structure in the final polish of the PS2 L3 lens. The fringe contours run smoothly through the center of the lens and much of the zonal structure (which was of minor importance with the PS1 L3 lens) has been eliminated.

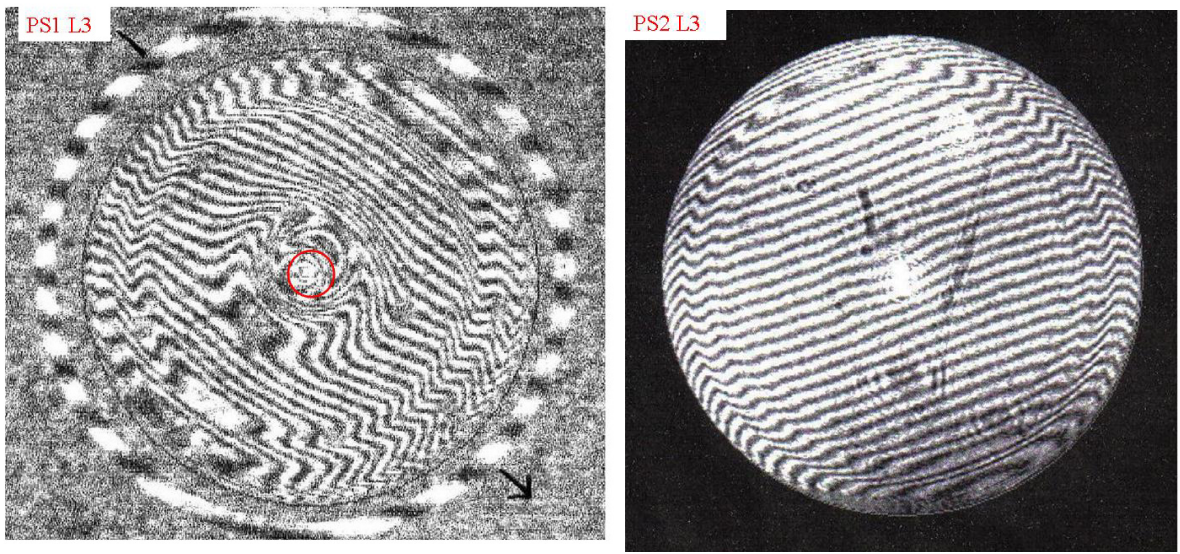

Figure 7. Interferometry of the PS1 and PS2 L3 corrector lenses

Figure 8 shows the interferometric measurements for the PS2 L2 corrector lens. No such data were available for the PS1 L2 lens. These measurements were made in a similar fashion to those shown earlier for the L3 lenses. Once again, the noisy bright spot near the center of this interferogram is due to out-of-focus higher-order diffraction from the hologram, but the fringes in these data do pass smoothly through the center of the interferogram. Like the PS2 L3 corrector lens polish, these data show no significant zonal structure and no significant polishing errors near the center of the lens. This result confirms that with the proper polishing techniques and without all of the initial lens generation problems, the L2 lens can be made to within the required specifications. In fact, the as-built lens polish for the L2 corrector lens considerably exceeds our specifications.

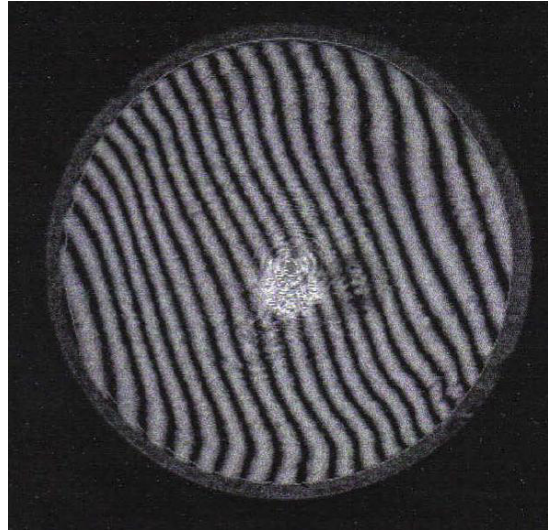

Figure 8. Interferometry of the PS2 L2 corrector lens.

Figure 9 shows ray trace calculations that illustrate the expected PS2 image quality based on the full aperture interferometry of all of the completed PS2 optics. Surface errors on both mirrors, all three corrector lenses, and the filters have been modeled and placed into an optical description of the telescope (NOADC-M-4.3.ZMX). The g filter was used in this illustration because it represents the most difficult passband in the system through which to achieve proper image quality. On the left are shown the focus curves for this realization of the design and on the right we show the RMS spot size expected as a function of focal plane position. The surface aberrations modeled cause the shift of the on-axis minimum focus towards the primary mirror by 40 μm. This shift is insignificant. More importantly, the minimum focus of the off-axis curves remains quite small, below 6 μm for all locations except for the extreme corners of the FOV where it raises to approximately 9 μm. All of these results are very encouraging and give us expectations for significant improvements in the performance of the PS2 optics over what we are currently working with in PS1. Note that a 10 μm RMS spot size is equivalent to a 0.36" FWHM.

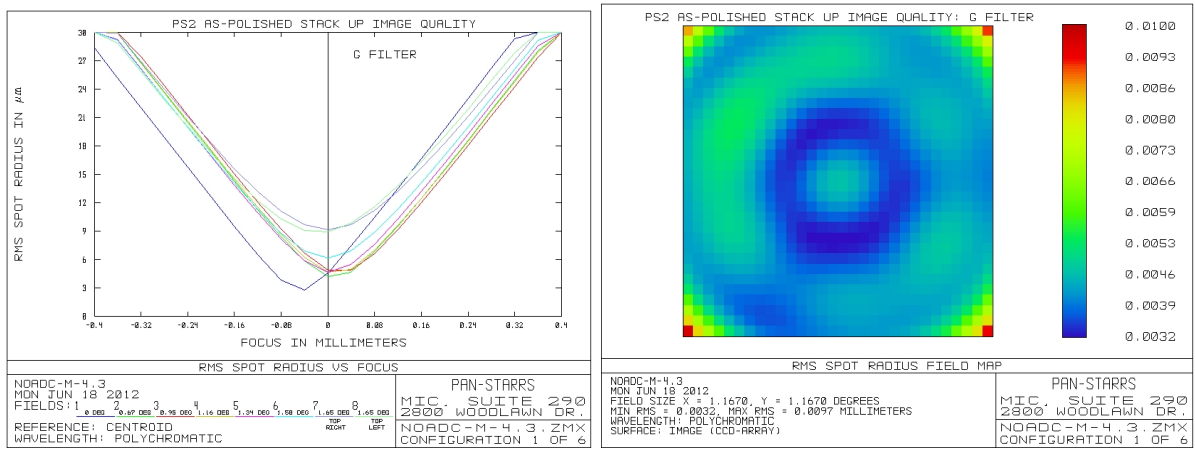

Figure 9. The expected as-polished image quality for PS2

## 4. CORRECTOR COATINGS

Ghosting in the PS1 telescope comes primarily from a reflection off of the detector surfaces themselves and then a return bounce off of the lower surface of the L1 corrector lens. The left side of Figure 10 shows the light path for the most prominent PS1 ghost images and the right side shows a typical example of a ghost in an r filter exposure caused by a bright star on the opposite side of the optical axis. The four cells shown in this figure represent a very small fraction of a single image (0.1% of a full image) and the so these ghosts are not a major limitation to the system performance. However, they are a significant annoyance to automated processing of the images so we would very much like to suppress them. The shapes and sizes of the ghosts vary with radius from the optical axis, becoming nearly linear at the FOV limits. The ghost intensity is passband dependent, being strongest in the g passband and getting rapidly weaker with wavelength until they are essentially gone in the z passband. This wavelength dependence is caused by both the CCD AR coatings and the AR coating on the bottom side of L1 (designated as L1b). A $5^{th}$ magnitude star in a g filter exposure will generate approximately 11 ADUs/second in a ghost footprint. The intensity through other filters is less than 2 ADUs/second for the same magnitude star.

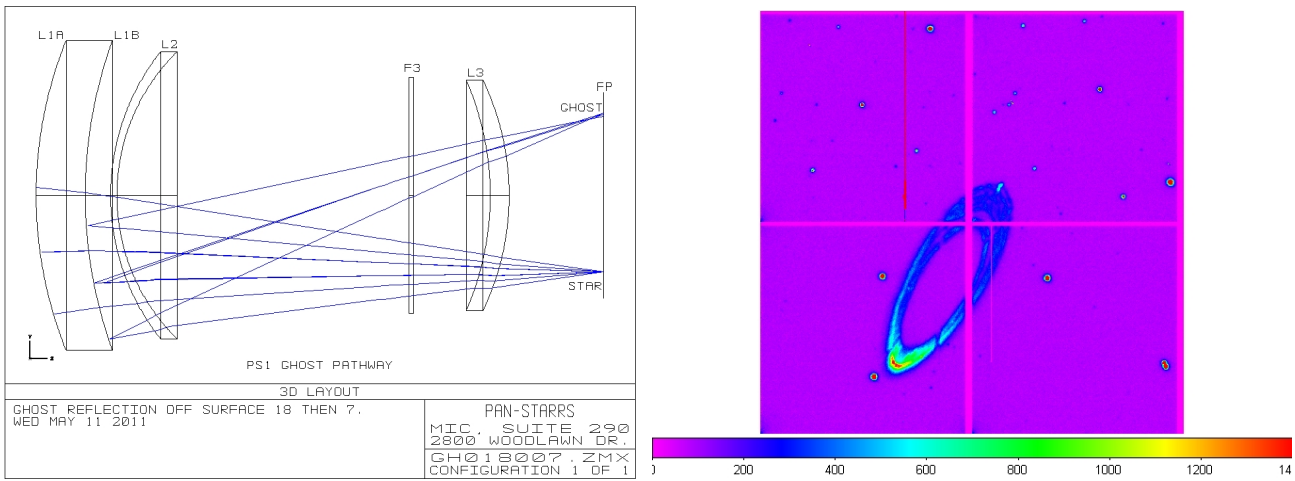

Figure 10. A typical CCD-L1b ghost in the PS1 telescope

To suppress these ghosts we have first pushed the highly reflective cutoff of the CCD AR coatings towards the blue, allowing us to optimize the design of the corrector lens AR coatings to minimize the ghost signal over the entire range of telescope operation. Figure 11 shows the as-built L1b AR coating spectral reflectance, measured at several radial positions across the lens. The coating uniformity across the lens is excellent, even in the g passband. The inset table gives an inverted figure of merit (FOM) in each passband; smaller numbers indicate better performance. This listed FOM in each passband should be proportional to the ghost brightness in a passband. The g band FOM is a factor of 12 below what we have for the coatings in PS1. This has been accomplished while keeping the FOM in all the other passbands lower everywhere but in the y filter, where the ghost brightness is expected to increase by a factor of 2. This is a very acceptable trade-off since the PS1 ghost brightness in the y filter is already 18 times fainter than the g filter ghosts.

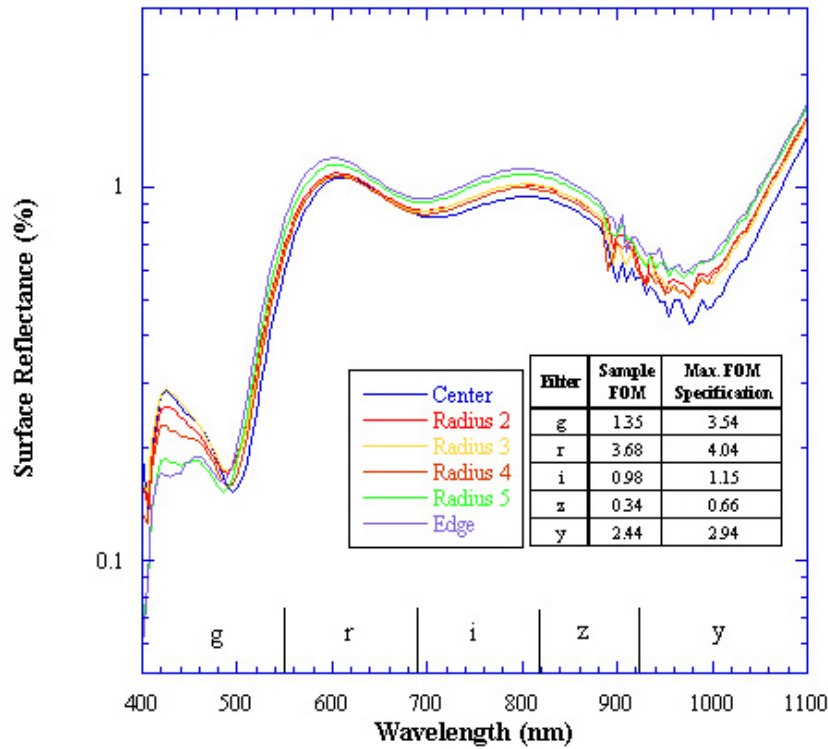

Figure 11. The PS2 L1b AR coating

## 5. REMOVAL OF PRIMARY MIRROR ASSEMBLY HEAT SOURCES

The PS1 telescope suffers from two major sources of heat: signal conditioners embedded inside the Primary Mirror Assembly (PMA) release a significant amount of heat as conductive print-through into the mirror cavity itself and the camera power supplies release a significant heat load underneath the PMA. It was not originally intended to position the camera power supplies directly under the PMA, but limitations to acceptable voltage drops at the camera head forced the decision to move these supplies onto the rotator assembly. Efforts to mitigate the camera heat loading by importing glycol cooling into the rotator instrumentation have largely been effective, but there is still residual camera heat that may be having a minor impact on the system. Here we will focus mostly on the telescope sources of heating.

Figure 12 and Figure 13 show the heat leakage into the PS1 PMA. On the left of each figure are shown visible images of the FLIR scenes seen on the right. The primary mirror, two of the telescope baffles, the primary mirror stop, the mirror covers, and the primary mirror lateral support actuators are seen in each image. In the visible images the dark rectangular boxes seen above and below the primary mirror are the pneumatic lateral support actuators. Being pneumatic, these actuators are not a source of heat. The PS1 signal conditioners are located underneath the white metal support structure below the lateral actuators near the top right corner of the PMA. The 2-3° increase in the metal below the pneumatic actuators is the heat print-through from the PMA signal conditioners.

The locations of most of the signal conditioners in the PS2 telescope have been moved to glycol-cooled electronics cabinets near the top and bottom of the PMA. The 36 axial actuators each contain electronics that dissipate 1.5 W of power each, but the sum power in these units is less than ½ of the power being dissipated inside the PS1 mirror cell. In addition, the azimuth and altitude motors themselves will be glycol cooled. We anticipate improvements in the dome seeing from these changes.

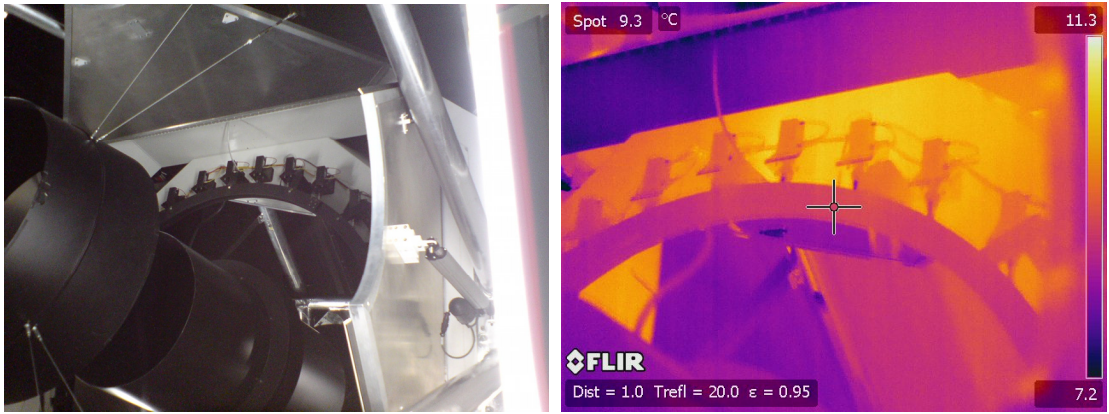

Figure 12. Heat leakage into the top of the PS1 mirror cell

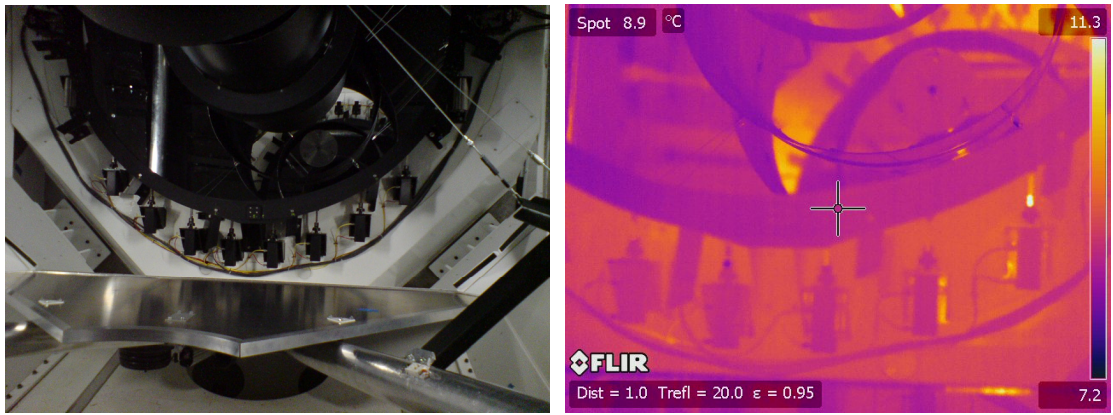

Figure 13. Heat leakage into the bottom of the PS1 mirror cell

## 6. INSTRUMENT CABLE WRAP CHANGES

The cable wrap in the PS1 telescope was designed and fabricated after the telescope itself was delivered. This was driven by uncertainties in the required cables for the camera itself. Initial concerns over the reliability of the filter changing mechanism drove the decisions to make the filter mechanism easily removable without disturbing the camera. These circumstances and decisions left no room for the development of a balanced cable wrap with the PS1 telescope. A single-sided wrap maintained the quick serviceability of the filter mechanism, but came at the cost of having the telescope properly balanced only at a limited range of instrument rotator positions. This places constant stress on the PS1 altitude motor and significantly limits its margin of torque for handling wind buffeting and slews. This also makes the telescope very difficult to balance properly. In retrospect, this was probably a poor design choice because the filter mechanism has proven to be a very reliable design. It has failed only once in its nearly continuous three years of operation and in a way that was quickly fixed without its removal from the telescope.

In contrast, the PS2 telescope design has had the benefit of known cable specifications prior to its development. Extra space in the design of the PMA and the rotator support structure was put in to accommodate both a dual-sided, balanced instrument rotator wrap and access for the quick removal of the filter mechanism. Figure 14 shows details of the PS2 instrument rotator design, including an expanded view of the cage structure supporting one side of the wrap. Owing to these changes, we anticipate better tracking performance under windy conditions and much greater ease of maintaining its proper balance compared to PS1.

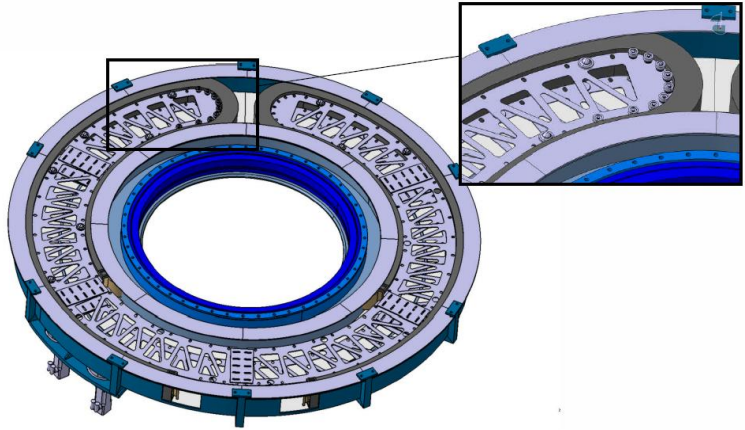

Figure 14. The PS2 Instrument Rotator (courtesy of AMOS)

## 7. M1 FIGURE CONTROL ACTUATORS

The PS1 telescope incorporates a primary mirror figure control system that consists of twelve differentially controllable force actuators near the perimeter of the mirror. The rest of the 36 axial M1 support points were ganged into three control loop sectors. This design is simple to control. It has been effective and has allowed us to achieve minimum distortions of the mirror over the full range of telescope positions, but the forces required of the figure control system to do this have been larger than anticipated and have taxed the ability of the system to induce the required support compensations. At elevations below ~30º, the system bottoms out and is no longer able to fully correct the mirror figure. Because of this, during the fabrication of the PS2 telescope, it was decided that it was cost effective to replace the "ganged" axial primary mirror actuators of the PS1 system with 36 completely independent axial actuators. Effectively, we have changed the design of the figure control system from 12 actuators in the outer ring of supports to 36 actuators in both inner and outer support rings. Figure 15 shows the PS2 primary mirror actuators prior to installation into the PMA.

This design change increases both the range of forces available for figure control and the modes of figure correction available. In particular, the 36 actuator figure control will be far more capable of inducing changes in the system which require differential radial variations in the support forces. It will be easier for us to induce wavefront changes in Zernike terms like focus. This is important because this optical system is very sensitive to variations in the primary mirror radius of curvature. This design change also gives us the potential for correcting more high order terms than we are capable of with the PS1 system.

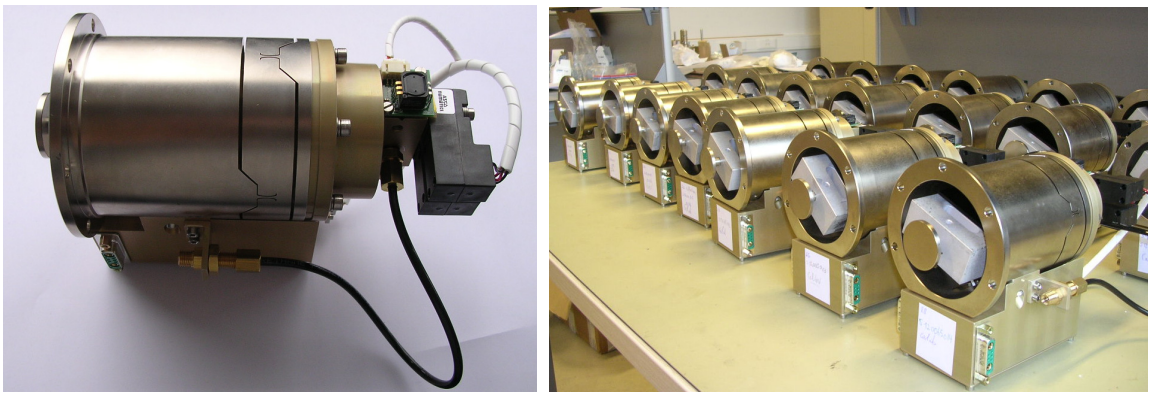

Figure 15. PS2 axial primary mirror actuators prior to installation into the PMA (photos courtesy of AMOS and Micromega Dynamics)

Another benefit of the PS2 actuators results from the fact that each actuator will have an independent load cell reading of the force being applied. While this induces complications in the system control and adds risk for parts failure, it has the benefit of direct monitoring of all forces placed on the mirror. This is an important benefit from the standpoint of system diagnostics.

## 8. IMPROVED BAFFLING

One of the most important lessons learned from the experience of building and operating the PS1 telescope has been how to properly minimize the scattered light in the system. Since the telescope passbands extend all the way out to 1 μm, it has been important to pay attention to surface reflectivities between 0.7 and 1.1 μm. Many surfaces that look black in the visible are quite reflective in this passband. Most of these deficiencies with the PS1 baffling have been corrected, but some require modifications to the corrector lens support structures and disassembly of the L1 and L2 optics and subsequent realignment of the optics to correct. The potting of the L1 and L2 lenses into their support cells represents a known contributor to the current scattered light background in the PS1 telescope. External baffles have been applied to reduce scattered light from these surfaces, but we believe that we can do better for PS2 with additional internal baffles.

Figure 16 shows a cross section of the PS2 L1 and L2 support structure. The L1 and L2 lenses are potted into their cells with an RTV compound that is designed to compensate for thermal expansion of the steel cells. Unfortunately, this material is white, highly reflective and its thermal properties are very important to the functioning of the lens support. To guarantee its reliable thermal properties we are reluctant to try additives to reduce its reflectivity. The L1 Top Baffle and the L2 Barrel Bottom Flange exist in the current telescope and are known to help considerably in reducing the scattered light from illumination of the RTV potting. However, moonlight from large angles (>20°) is still capable of making it past the 3 main system baffles and illuminating this potting material, both directly and through the glass of the corrector lenses. For PS2 we plan to add two features to the system that we believe will offer improvements in scattered light rejection. First, we are placing a layer of dark paint between the RTV coating and the lens glass. Adhesion tests have shown that this is acceptable. Second, we are installing the two intervening baffles to the PS2 system labeled "PS2 L1 Cell Base" and "PS2 L2 Top Baffle" in Figure 16. When the PS1 system becomes available for engineering upgrades we plan on retrofitting its optics with the new baffles, but it's unlikely that we will be able to re-do the RTV potting to add the paint layer.

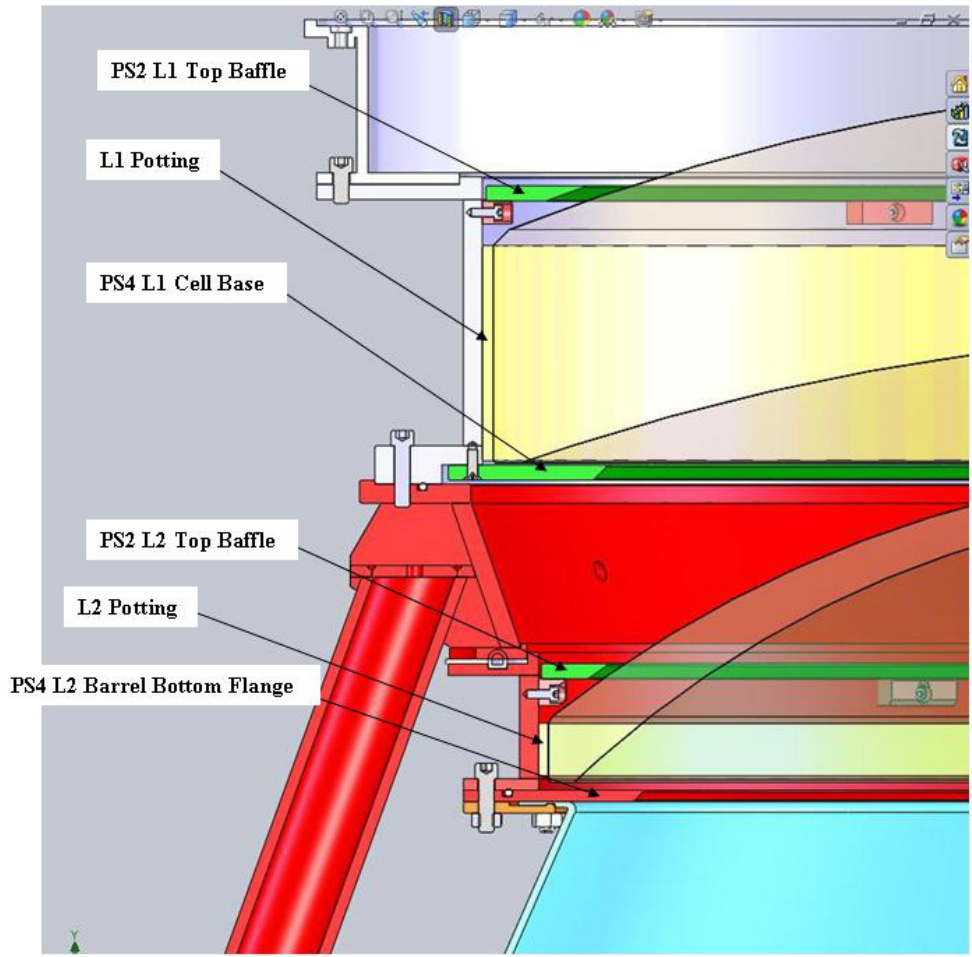

Figure 16. The PS2 UCC baffles

## 9. SUMMARY

Fabrication and design changes have been made to the PS1 configuration that offer significant promise for the performance of the PS2 telescope. It is our expectation that these changes will result in improved image quality, improved performance in windy conditions, decreased ghosting, and improved scattered light rejection.